\documentclass[twocolumn,english]{revtex4-1}
\usepackage[T1]{fontenc}
\usepackage[latin9]{inputenc}
\usepackage{float}
\usepackage{textcomp}
\usepackage{amsmath}
\usepackage{amssymb}
\usepackage{graphicx}

\makeatletter

\providecommand{\tabularnewline}{\\}

\makeatother

\usepackage{babel}
\begin{document}
\title{On the potential of probing the neutron star composition in accreting
X-ray binaries}
\author{Kaiser Arf and Kai Schwenzer }
\address{Istanbul University, Science Faculty, Department of Astronomy and
Space Sciences, Beyazit, 34119, Istanbul, Turkey}
\begin{abstract}
Transiently accreting Low Mass X-Ray Binaries have the potential to
probe the core composition of their neutron stars via deep crustal
heating caused by nuclear reactions. We statistically assess this
deep crustal heating scenario, taking into account the various microphysical
and astrophysical uncertainties. We find that despite the sizable
uncertainties there is the chance to discriminate different compositional
scenarios. Several observed sources statistically challenge a minimal
hadronic matter composition, where cooling proceeds exclusively via
slow modified Urca reactions. Considering here two exemplary extended
uniform compositions, namely ultra-dense hadronic matter with direct
Urca emission and ungapped quark matter, we find that they are even
within uncertainties distinguishable. We show that although exotic
forms of matter are generally only expected in an inner core, which
could in principle have any size, sufficiently large astrophysical
data sets nonetheless have the potential to statistically discriminate
compositional scenarios, in particular when further mass measurements
become available.
\end{abstract}
\maketitle

\section{Introduction}

The observed Low Mass X-ray Binaries (LMXBs) \citep{Liu:2007ts} are
thought to be neutron stars \citep{1983bhwd.book.....S,Haensel2007}
in the later stage of their recycling process \citep{Alpar:1982Natur.300..728A},
accreting from a low mass companion via Roche lobe overflow \citep{potekhin2010physics}.
Many of them are transiently accreting systems and show strong, and
typically short, outbursts followed by longer periods of quiescence,
where the luminosity is significantly smaller. While both the peak
luminosity and the quiescence period can vary between outbursts as
well as between different sources, over long time intervals these
sources are by energy conservation expected to reach a steady state
where the accretion heating balances the cooling due to both bulk
neutrino and surface photon emission \citep{Yakovlev2004,Wijnands2017,Potekhin2019}. 

The interior of the star is strongly thermally insulated from its
surface by the outermost layers of the crust where both the density
and the thermal conductivity decrease dramatically, so that nearly
all of the gravitational potential energy deposited during outburst
is quickly radiated from the surface layer, having a small heat capacity
\citep{Chamel2008,beznogov2021heat,Beznogov2017}. Yet, over time
material is buried under a growing layer of subsequently accreted
material and eventually transforms to successively heavier nuclei
as the density and pressure on it increases. It has been shown that
the nuclear energy from pycno-nuclear reactions deep in the crust,
fusing lighter into heavier nuclei, can produce a significant amount
of heat \citep{1990A&A...227..431H,Potekhin1997,Haensel2003,Haensel2008,Fantina:2018yad}.
This deep crustal heating \citep{Brown1998} directly heats the core
of the star and can therefore inform us about the cooling process(es)
\citep{Yakovlev2004}, which strongly depend on the composition. This
long term thermal state should be reflected after a sufficiently long
time in quiescence when the thermal profile in the crust relaxed back
to its steady state profile, while the average heating power can be
obtained from the luminosities in and time intervals between outbursts
\citep{Potekhin2019}.

Since the long-term steady state of the system is completely determined
by energy conservation, this provides for a given star composition
a direct analytic relation between the observable thermal quiescence
luminosity and average accretion rate, that can be compared to astrophysical
data. However, this relation involves many microscopic and macroscopic
parameters some of which involve significant uncertainties. The largest
uncertainty stems from the surface composition \citep{Brown:2002rf,beznogov2021heat}
and we consider the two extreme cases of a fully catalyzed \citep{Gudmundsson1982}
and a fully accreted envelope \citep{Potekhin1997} separately. We
estimate and take into account the other uncertainties and perform
a statistical analysis to determine how compatible a given compositional
scenario is with the astrophysical data for a given source. We quantify
this by corresponding probabilities and find that even taking into
account these various uncertainties some of these probabilities are
remarkably small, so that the corresponding compositional scenario
can be statistically ruled out. Taking into account that the effect
of superfluidity is still rather uncertain, we use the minimal cooling
of $npe\mu$-matter with modified Urca neutrino processes \citep{Friman1979}
as a baseline, and find that there are several sources that challenge
this minimal scenario. This requires enhanced cooling mechanisms and
we consider here the extreme case of direct Urca processes in both
hadronic \citep{Lattimer1994} and quark matter \citep{Iwamoto1980}.

\section{Long-term thermal steady state\label{sec:Steady-state}}

In compact stars the inner energy (and its change) is tiny compared
to their total energy and therefore the thermal evolution can be considered
within the background structure of a static cold star \citep{Thorne1967hea3.conf..259T,Thorne1977ApJ...212..825T}.
The metric of a spherically symmetric mass configuration is described
by the Schwarzschild metric 

\[
ds^{2}=e^{2\phi(r)}dt^{2}-\left(1-\frac{2Gm\!\left(r\right)}{r}\right)^{-1}dr^{2}-r^{2}d\Omega^{2}
\]
in terms of two metric functions: $m\!\left(r\right)$ the mass within
radius $r$, and $\phi\!\left(r\right)$ which is in the non-relativistic
limit related to the Newtonian gravitational potential. A static compact
star is then completely described by the density distribution and
composition within the star and is obtained from a solution of the
TOV equation \citep{Oppenheimer1939,Tolman1939}. At the surface and
outside of the star, where $m\!\left(r\right)$ is constant $m\!\left(r\geq R\right)=M$,
the curvature is completely described by the potential $\phi(r\geq R)=\frac{1}{2}\log\!\left(1-\frac{2GM}{r}\right)$.

In the most outer part of the crust of a neutron star the density
and correspondingly also the thermal conductivity drops dramatically,
effectively presenting an insulating layer. The temperature $T_{b}$
at the inner boundary of this insulating layer can therefore strongly
differ from temperature the $T_{s}$ at the surface \citep{yakovlev2004neutron},
which due to the small thickness of the insulating layer is roughly
obtained at the radius $R$. The global thermal evolution of a an
accreting star is determined by energy conservation and takes the
form \citep{Glen1980,Yakovlev:2002ti}: 

\begin{equation}
C_{V}(T_{b})\frac{dT_{b}}{dt}=\text{\ensuremath{P_{h}^{\infty}}}-L_{\nu}^{\infty}\!\left(T_{b}\right)-\text{\ensuremath{L_{\gamma}^{\infty}}}\!\left(T_{s}\right)
\end{equation}
where $\text{\ensuremath{P_{h}^{\infty}}}$, $L_{\nu}^{\infty}$ and
$L_{\gamma}^{\infty}$ are the heating power, the neutrino luminosity
from the bulk and the surface photon luminosity, all considered at
a reference frame at infinity.

When averaging over many outburst/quiescence cycles, the star should
eventually reach a steady-state \citep{Yakovlev:2002ti,Yakovlev2004,Potekhin2020}
where the temperature does not change anymore and correspondingly
the heating roughly balances the cooling 

\begin{equation}
\text{\ensuremath{P_{h}^{\infty}}}\approx L_{\nu}^{\infty}+\text{\ensuremath{L_{\gamma}^{\infty}}}\label{eq:steady-state}
\end{equation}
Here we consider the strong heating due to the accretion, neglecting
other potential heating mechanisms, e.g. roto-chemical heating \citep{Reisenegger:1994be}
or dissipative damping of unstable oscillation modes \citep{Andersson:2000mf,Alford:2010fd}.
In particular r-mode heating could be sizable if the r-mode amplitude
is large. However, by now there are strict bounds on the r-mode amplitude
in millisecond pulsars \citep{Boztepe:2019qmm} that are more than
an order of magnitude below those from LMXBs and since the r-mode
saturation mechanism should at most be weakly temperature dependent
also r-mode heating should indeed be minor in LMXBs. LMXBs, being
very old, have small dipole magnetic fields so that their potential
impact on the heating can likewise be neglected.

While the surface heating from the accretion during outburst is, owing
to the insulating layer and the small heat capacity of the outermost
part of the crust, nearly immediately radiated off, only the heating
due to pycno-nuclear reactions deep in the crust can actually heat
the interior. This deep crustal heating is determined by the average
amount of energy per baryon $Q$ deposited deep in the crust \citep{Haensel2003,haensel2008models,Fantina:2018yad},
and the heating power as seen from infinity reads \footnote{The heating is localized at the boundary where the heat balance can
be established. Correspondingly the heating power at infinity is a
mere definition arising when rewriting the heat balance equation in
terms of luminosities at infinity. The linear dependence on the metric
factor compared to the luminosities below that are quadratic in it,
stems from the fact that the energy gained when matter is falling
into the gravitational potential well does not contribute to deep
crustal heating but is directly radiated off at the surface. }

\begin{equation}
P_{h}^{\infty}\equiv\frac{Q}{m_{N}}\left\langle \dot{M}\right\rangle e^{\phi\!\left(R\right)}
\end{equation}
where $e^{\phi\!\left(R\right)}$ is the spatial component of the
Schwarzschild metric at the surface of the star, $m_{N}$ the nucleon
mass and $\left\langle \dot{M}\right\rangle $ is the average mass
accretion rate. 

The neutrino luminosity depends strongly on the core composition and
the temperature. The core temperature of a neutron star is not directly
observable but sufficiently long after outburst the crust, which was
pushed out of equilibrium by the accretion heating, relaxes back to
the quasi-equilibrium thermal profile and there is a unique dependence
between the temperature $T_{b}$ below and the surface temperature
$T_{s}$ above the insulating layer that takes the form \citep{Gudmundsson1982}

\begin{equation}
\frac{T_{s}^{4}}{g_{s}}=f\!\left(T_{b}\right)\approx cT_{b}^{4\gamma}\label{eq:core-surface-relation}
\end{equation}
where $g_{s}$ is the surface gravitational field $g_{s}=e^{-\phi\left(R\right)}GM/R^{2}$
in terms of the temporal metric factor $e^{\phi}$. The precise function
$f$ in eq. (\ref{eq:core-surface-relation}) is obtained by solving
the heat-conduction throughout the insulating layer and over the relevant
temperature range it is to good accuracy described by a simple power
law form \citep{Gudmundsson1982,Potekhin:1997mn}. The precise relation,
described by the parameters $c$ and $\gamma$, depends on the composition
of the crust.

Owing mainly to the heat transport by light and merely electromagnetically
interacting electrons, the thermal conductivity below the insulating
layer is very large. Thorne \citep{Thorne1967hea3.conf..259T,Thorne1977ApJ...212..825T}
and Glen \citep{Glen1980} studied thermal transport in the curved
spacetime inside of a neutron star and found that owing to energy
conservation even in the limit of an infinitely large thermal conductivity
the star is not isothermal, but rather the total particle energy (kinetic
and potential) is constant resulting in $T\left(r\right)e^{\phi\left(r\right)}=\mathrm{const}$.
Using this relation, the temperature anywhere inside the star is given
in terms of the temperature $T_{b}$ at the inner boundary of the
insulating layer, which is approximately located at the radius $R$
of the star, via

\begin{equation}
T(r)=T_{b}\frac{e^{\phi(R)}}{e^{\phi(r)}}\label{eq:17_local_boundaryT}
\end{equation}
The local neutrino emission due to a process $i$ is described by
the neutrino emissivity $\epsilon_{i}(r)$ which in general has the
following dependence on the local temperature \citep{Alford:2013pma}

\begin{equation}
\epsilon_{i}(r)=\hat{\epsilon}_{i}\left(n(r),x_{s}\left(n(r)\right),...\right)T(r)^{\theta_{i}}\lambda\left(r\right)^{\Theta_{i}}\label{eq:emissivity-parametrization}
\end{equation}
where $\theta_{i}$ denotes the power law exponent characterizing
the mechanism, $\tilde{\epsilon}_{i}$ is a temperature-independent
pre-factor and the last factor describes logarithmic non-Fermi liquid
corrections in certain phases of dense quark matter \citep{Schafer:2004jp},
which we will neglect in this work (i.e. here $\Theta=0$). The pre-factor
generally depends on other local quantities like the overall baryon
density $n$ as well as particle fractions $x_{s}$ of the different
species $s$, which in equilibrium are functions of the baryon density.
Taking into account that a neutron star is a strongly degenerate system
where $T/\mu\lesssim10^{-4}$ in particular the exponent $\theta$,
ranging from slow processes ($\theta=8$) to fast processes ($\theta=6$),
has a dramatic impact on the cooling \citep{Yakovlev:2000jp}. It
is determined by qualitative aspects of the low energy degrees of
freedom and their interactions and alters the luminosity by many orders
of magnitude for different phases of dense matter, potentially allowing
us a clear discrimination despite the various micro- and macro-physical
uncertainties.

 The total luminosity at infinity is then obtained by integrating
the neutrino emissivity over the entire star, taking into account
the spatial volume factor $\sqrt{\det\!\left(g_{ij}\right)}=\left(1-2Gm\!\left(r\right)/r\right)^{-1/2}$
and the appropriate red-shifting,
\begin{align}
 & L_{\nu,i}^{\infty}=\int d^{3}r\sqrt{\det\!\left(g_{ij}\right)}\label{eq:neutrino_luminosity}\\
 & \qquad\times\tilde{\epsilon}_{i}\!\left(n\!(r),x_{s}\!\left(n\!(r)\right),...\right)\left(\!T_{b}\frac{e^{\phi\!\left(R\right)}}{e^{\phi\!\left(r\right)}}\!\right)^{\!\theta_{i}}\!e^{2\phi\left(r\right)}\nonumber 
\end{align}
This expression can be be written in the form

\begin{equation}
L_{\nu,i}^{\infty}=\frac{M}{n_{0}m_{N}}\hat{\epsilon}_{i}\left(n_{0},x_{s}\!\left(n_{0}\right),...\right)T_{b}^{\theta_{i}}\tilde{L}_{i}\label{eq:luminosity-parametrization}
\end{equation}
in terms of the luminosity of a hypothetical uniform star of mass
$M$ in the absence of gravity---i.e. literally a dramatically scaled
up atomic nucleus, with matter at saturation density $n_{0}$ \footnote{Even though such a configuration could in principle be stable in the
absence of gravity, it would require gravity to form it.}, having a Euclidean volume $V=M/\left(n_{0}m_{N}\right)$, where
$m_{N}$ is the nucleon mass ---and a dimensionless constant factor
$\tilde{L_{i}}$ that takes into account all effects of gravity, i.e.
the detailed spacetime curvature, density distribution and local composition
within the star

\begin{align}
\tilde{L}_{i} & \equiv\frac{4\pi n_{0}m_{N}}{M}\!\int_{0}^{R}\negthickspace dr\,r^{2}\left(1\!-\!\frac{2Gm\!\left(r\right)}{r}\right)^{\!-1/2}\label{eq:luminosity-constant}\\
 & \qquad\qquad\times\frac{\hat{\epsilon}_{i}\!\left(n(r),x_{s}\!\left(n(r)\right),...\right)}{\hat{\epsilon}_{i}\!\left(n_{0},x_{s}\!\left(n_{0}\right),...\right)}\!\left(\frac{e^{\phi\left(R\right)}\!}{e^{\phi\left(r\right)}}\right)^{\!\theta_{i}}e^{2\phi\left(r\right)}\nonumber 
\end{align}
This dimensionless constant absorbs the entire dependence on the
poorly known equation of state. All factors in the emissivity ratio
that are not radius-dependent cancel in eq. (\ref{eq:luminosity-constant})
and this constant can be numerically computed for different star configurations
based on different equations of state. Interestingly, we find that
$\tilde{L}$ is close to $1$ for a $1.4\,M_{\text{\ensuremath{\odot}}}$
star with an APR equation of state \citep{Akmal:1998cf} cooling via
Urca processes \citep{Friman1979,lattimer1991direct}---i.e. the
total effect of gravity is actually not that big, and this only strongly
changes as one goes close to the maximum mass configuration.

The neutrino luminosity for known emission mechanisms is strongly
temperature-dependent ($\theta_{i}\gtrsim6$) and therefore it largely
dominates the photon surface emission $L_{\gamma}\sim T_{s}^{4}$
sufficiently above the temperature $T_{e}$, where they are of equal
size, and approximately alone balances the heating in eq. (\ref{eq:steady-state}),
i.e. $L_{\nu,i}^{\infty}\!\left(T_{b}\right)\approx\text{\ensuremath{L_{h}^{\infty}}}$,
which determines the steady state temperature below the insulating
layer. Using eq. (\ref{eq:core-surface-relation}) this can be related
to the surface temperature outside of the insulating layer, and using
the Stefan-Boltzmann law finally to the effective photon luminosity
in quiescence red-shifted to infinity, which takes for $T$ sufficiently
above $T_{e}$ the explicit analytic form

\begin{align}
 & L_{\gamma}^{\infty}=4\pi\sigma R^{2}T_{s}^{4}e^{2\phi\left(R\right)}=4\pi\sigma R^{2}g_{s}cT_{b}^{4\gamma}e^{2\phi\left(R\right)}\label{eq:luminosity-mass-accretion-relation}\\
 & \approx4\pi\sigma cGM^{1-\frac{4\gamma}{\theta_{i}}}e^{\left(1+\frac{4\gamma}{\theta_{i}}\right)\phi\left(R\right)}\negmedspace\left(\!\frac{n_{0}Q\left\langle \dot{M}\right\rangle }{\hat{\epsilon}_{i}\!\left(n_{0},x_{s}\!\left(n_{0}\right),...\right)\tilde{L}_{i}}\!\right)^{\negmedspace\frac{4\gamma}{\theta_{i}}}\nonumber 
\end{align}
where $\sigma$ is the Stefan-Boltzmann constant. Note that in eq.
(\ref{eq:luminosity-mass-accretion-relation}) $\theta_{i}>4$ and
$\gamma<1$ so that this result is fortunately rather insensitive
to the various parameters describing the process, which can have sizable
uncertainties. In particular the leading power-law dependence on the
radius cancels out and leaves only the indirect dependence via the
metric factor. Yet, eq. (\ref{eq:luminosity-mass-accretion-relation})
sensitively depends on the power law exponents $\theta_{i}$ and $\gamma$
characterizing the neutrino emission in qualitatively different forms
of dense matter in the interior of the star, as well as the crust
composition, respectively.

Averaging over several outbursts the average accretion rate in turn
can be obtained from the average outburst luminosity and the average
recurrence time, taking into account that the gravitational potential
energy of the accreted mass is deposited as heat on the surface and
(owing to the small heat capacity of the envelope) immediately radiated
off during the accretion phase. The accretion should happen dominantly
at the magnetic poles of the star creating hot spots. Yet the accreted
material and the heat is eventually distributed over the surface of
the star. As long as the surface luminosity significantly exceeds
the luminosity of the hot spot (which is typically the case due to
its small size), this complication can be neglected. 

If in contrast the temperature is sufficiently below $T_{e}$ the
photon luminosity strongly dominates neutrino cooling so that the
photon luminosity fulfills the simpler relation 
\[
L_{\gamma}^{\infty}\approx\text{\ensuremath{P_{h}^{\infty}=\frac{Q}{m_{N}}\left\langle \dot{M}\right\rangle e^{\phi\!\left(R\right)}}}
\]
Correspondingly the deep crustal heating scenario cannot probe the
interior composition of the star in cold sources and the detailed
transition point $T_{e}$ depends on the emission mechanism. In between
these two regimes there is a (typically narrow) transition region,
where both cooling mechanisms are relevant, that can be studied by
a numerical solution of eq. (\ref{eq:steady-state}). Here we will
concentrate on the asymptotic regimes, since only the explicit analytic
expressions allow us a systematic assessment of the various uncertainties
involved in the analysis.

Finally in a star with fast (direct Urca) cooling $L_{\nu f}^{\infty}$
in an inner core of mass $M_{c}$ and slow (modified Urca) cooling
$L_{\nu f}^{\infty}$ in the surrounding mantle, neglecting surface
emission, the steady state equation can be written in the terms of
the mass fraction

\begin{equation}
\text{\ensuremath{P_{h}^{\infty}}}\approx\left(1-\frac{M_{c}}{M}\right)L_{\nu s}^{\infty}+\frac{M_{c}}{M}L_{\nu f}^{\infty}\label{eq:hybrid-steady-state}
\end{equation}
which gives for the required mass fraction at a given temperature
$T_{b}$

\begin{equation}
\frac{M_{c}}{M}\approx\frac{P_{h}^{\infty}-L_{\nu s}^{\infty}\!\left(T_{b}\right)}{L_{\nu f}^{\infty}\!\left(T_{b}\right)-L_{\nu s}^{\infty}\!\left(T_{b}\right)}\label{eq:mass-fraction}
\end{equation}

\section{Star composition}

\subsubsection{Neutron star envelope $T_{b}-T_{s}$ relation }

The composition of the envelope of the neutron star has a strong impact
on the insulation and its composition is altered by the accreted matter
\citep{Haensel2007}. The envelope composition thereby introduces
a significant uncertainty when trying to determine the interior composition
of the star from astrophysical data. Here we consider two extreme
cases: a pure iron (non-accreted) envelope \citep{Gudmundsson1982}
and a light element (fully accreted) envelope \citep{Potekhin1997}.
The envelopes of actual accreting sources should be somewhere in between
these two limits, but we refrain from considering such intermediate
cases since they depend on model assumptions on the composition of
the crust. The $T_{b}-T_{s}$ relation for a non-accreted iron envelope
was given in \citep{Gudmundsson1982}, were a good fit to numerical
heat transport simulations throughout the heat blanket takes the form
\begin{equation}
T_{s6}=\left(0.1288^{-\frac{1}{0.455}}g_{s14}\right)^{\frac{1}{4}}T_{b9}^{\frac{1}{1.82}}\label{eq:non-accreted-envelope}
\end{equation}
where $T_{b9}$ is the normalized boundary temperature $T_{b9}=T_{b}/10^{9}\,\mathrm{K}$,
$T_{s6}$ is the normalized surface temperature $T_{s6}=T_{s}/10^{6}\,\mathrm{K}$
and $g_{s14}\equiv g_{s}/10^{14}\text{cm/}\text{s}^{\text{2}}$is
the normalized surface gravitational acceleration of neutron stars.
Similarly the surface temperature relation for a fully accreted envelope
was been given for boundary temperatures $T_{b}\leq10^{8}\,\mathrm{K}$
as \citep{Potekhin1997} 
\begin{equation}
T_{s6}=\left(18.1^{2.42}g_{s14}\right)^{\frac{1}{4}}T_{b9}^{\frac{2.42}{4}}\label{eq:accreted-envelope}
\end{equation}

\subsubsection{Basic neutrino emission processes}

The local, microscopic quantity that can distinguish different compositions
in the neutron star interior is the neutrino emissivity \citep{Yakovlev:2000jp},
dominating the heat loss in sufficiently hot sources. In ungapped,
uniform phases of matter it is generally Urca processes that dominate
the neutrino emission. The simplest composition is hadronic $npe\mu$-matter
where direct Urca processes are forbidden by momentum conservation
at sufficiently low density. In this case only modified processes
\citep{Friman1979} are possible where another collision with a bystander
particle via the strong interaction is necessary to provide the required
momentum

\begin{align}
\left(n,p\right)+p+e^{-} & \longrightarrow\left(n,p\right)+n+\nu_{e}\\
\left(n,p\right)+n & \longrightarrow\left(n,p\right)+p+e^{-}+\bar{\nu}_{e}
\end{align}
and these processes are correspondingly strongly suppressed. The star
is in chemical equilibrium with respect to these processes and moreover
has to be locally charge neutral. The expression for the modified
URCA neutrino emissivity in the pion-exchange approximation valid
for the long-range part of the nuclear force was derived in \citep{Friman1979}
and an approximation for its short-range component was considered.
The long range part is obtained from a chiral expansion and takes
the simple form (in natural units $\hbar=c=1$)

\begin{equation}
\epsilon_{\mathrm{mU}}\!=\!\frac{11513}{60480}\frac{G_{F}^{2}g_{A}^{2}}{2\pi}\alpha_{\mathrm{Urca}}\beta_{\mathrm{Urca}}\!\left(\!\frac{f}{m_{\pi}}\!\right)^{4}\!\left(m_{n}^{*}\right)^{3}\!m_{p}^{*}p_{Fe}\!\left(k_{B}T\right)^{8}\label{eq:modified-Urca-emissivity}
\end{equation}
from which one can read off $\hat{\epsilon}_{\mathrm{mU}}$ and $\theta_{\mathrm{mU}}=8$,
which characterizes a \emph{slow} process. Here $G_{F}$ is the Fermi
coupling, $g_{A}$ is the axial coupling and $m^{*}=p_{F}/v_{F}$
are effective masses,The parameter $\alpha_{\mathrm{Urca}}$ is a
correction factor including both a mild non-power-law density dependence
as well as off-shell effects, that were neglected in \citep{Friman1979},
but turned out to be sizable and were taken into account in \citep{Shternin:2018dcn}.
The contribution from short-range correlations is due to the present
impossibility to solve strongly coupled QCD at high density \citep{Brambilla:2014jmp}
still rather model dependent and in particular could also have a somewhat
different density dependence than eq. (\ref{eq:modified-Urca-emissivity}).
To take into account these uncertainties without having to make particular
assumptions on the detailed form of the emissivity, we added another
correction factor $\beta_{\mathrm{Urca}}$ in eq. (\ref{eq:modified-Urca-emissivity})
that reflects this uncertainty and which will be chosen below in order
that the expected physical curve lies within the corresponding enlarged
error band. The arising electron Fermi momentum is given in terms
of the baryon density and the electron fraction via $\mu_{e}\approx p_{Fe}=(3\pi^{2}x_{e})^{\frac{1}{3}}n^{\frac{1}{3}}$.
There are even slower processes (like bremsstrahlung), but since modified
Urca will be there as long as the nucleons are not entirely paired
these are typically negligible.

At sufficiently high density where the proton fractions exceeds a
critical value $x_{pc}$ somewhere between $11-15\%,$ depending on
the muon concentration, the nucleonic direct Urca channel 

\begin{align}
p+e^{-} & \longrightarrow n+\nu_{e}\label{eq:45_elec_cap}\\
n & \longrightarrow p+e^{-}+\bar{\nu}_{e}\label{eq:beta_decay}
\end{align}
gradually opens up over a rather narrow density interval where the
Urca emissivity ramps up by many orders of magnitude \citep{Shternin:2018dcn}
(while the corresponding direct muon process opens only at even higher
proton fractions and is therefore mostly negligible). This transition
can approximately be described by a step function $\Theta$, and and
its emissivity is given by \citep{Lattimer1994}

\begin{equation}
\epsilon_{\mathrm{dU}}=\frac{457\pi}{10080}G_{F}^{2}\cos^{2}\theta_{c}\left(1+3g_{A}^{2}\right)m_{n}^{*}m_{p}^{*}\mu_{e}\left(k_{B}T\right)^{6}\Theta\label{eq:hadronic-Urca-emissivity}
\end{equation}
which represents a \emph{fast} process with $\theta_{\mathrm{dU}}=6$.
Taking into account that in neutron stars $T_{b}/p_{F}=O\left(10^{-4}\right)$,
the direct rate is parametrically enhanced by many orders of magnitude.
Yet, relying merely on weak interactions this process is less uncertain
than the modified version. If the central density is above the critical
density for direct Urca reactions there will be a core within the
neutron star that emits with the much higher emissivity eq. (\ref{eq:hadronic-Urca-emissivity}),
whereas the rest of the star emits with the much lower emissivity
eq. (\ref{eq:modified-Urca-emissivity}) and will be negligible if
the core is sufficiently large.

Owing to the asymptotic freedom of QCD, at high density deconfined
3-flavor quark matter ($u$, $d$, $s$) could exist in the star.
In this case the analogous Urca processes at the quark level are relevant

\begin{align}
u+e & \longrightarrow d+\nu_{e}\\
d & \longrightarrow u+e+\bar{\nu}_{e}
\end{align}
The neutrino emissivity for quark Urca processes identically vanishes
for massless, non-interacting fermions \citep{Iwamoto1980,Alford:2011df}.
According to chiral symmetry restoration in deconfined quark matter
the quark masses are expected to be small, but since the interactions
are still sizable, the leading correction stems from radiative corrections
to the quark self-energy and the emissivity gives to leading order
in resummed perturbation theory \citep{Iwamoto1980}, 

\paragraph{
\begin{equation}
\epsilon_{\mathrm{qU}}=\frac{914}{315}G_{F}^{2}\cos^{2}\theta_{c}\text{\ensuremath{\alpha_{s}}}p_{Fd}p_{Fu}p_{Fe}\left(k_{B}T\right)^{6}\label{eq:quark-Urca-emissivity}
\end{equation}
}

which is likewise a fast process $\theta_{\mathrm{qU}}=6$, but since
the direct Urca rate vanishes for free ultra-relativistic particles
and quarks are much closer to that limit than massive hadrons the
prefactor of the quark emissivity is parametrically smaller than that
of the hadronic process. The analogous neutrino processes involving
a strange quark are suppressed in the Cabbibo angle $\theta_{c}$
and can to leading order be neglected. Taking into account that symmetric
3-flavor quark matter is automatically charge neutral, the electron
Fermi momentum is determined by the mass imbalance and when neglecting
the masses of the light quarks effectively by the strange quark mass
$p_{Fe}\approx m_{s}^{2}/\left(4p_{Fq}\right)$. Since eq. (\ref{eq:quark-Urca-emissivity})
is a result of resummed perturbation theory it could receive sizable
corrections in the still rather strongly coupled regime realized in
neutron star matter. This dependence on non-perturbative effects is
also reflected in the size of the strong coupling $\alpha_{s}$ which
is already rather uncertain in the vacuum \citep{Deur:2023dzc}, and
even more so in dense matter. Instead of introducing another multiplicative
parameter that takes into account that non-perturbative effects will
change this result, we rather consider the strong coupling in eq.
(\ref{eq:quark-Urca-emissivity}) as an effective parameter and the
uncertainty range we use for it below will include the combined estimated
impact of non-perturbative effects on the neutrino emission. 

Interacting quark matter can also be in a color superconducting state
\citep{Alford:2007xm}. Even in this case there are typically ungapped
quark flavors which dominate the neutrino emission, and sufficiently
below the critical temperature of the superconductor the emissivity
is in uniform phases effectively simply given by eq. (\ref{eq:quark-Urca-emissivity})
times a suppression factor reflecting the reduced number of ungapped
flavors. The exception is the CFL phase \citep{Alford:2007xm} where
all quark flavors pair in a particular symmetric way. As mentioned,
if the quarks interact via unscreened gluons, which is the case in
certain quark phases, the emissivity involves a non-Fermi liquid enhancement
\citep{Schafer:2004jp} which introduces an additional logarithmic
temperature dependence that we will neglect in this work. 

The fact that the simplest compositional scenario, namely hadronic
$npe\mu$-matter, can feature both very slow as well as the fastest
known processes, and that, depending on the equation of state, the
fast component can be realized in a core of varying size, shows that
it is inherently hard to observationally confirm other compositional
scenarios. Nonetheless there is a unique equation of state in nature
and therefore a very narrow mass range where the direct Urca channel
opens up. Therefore, with sufficient statistics and additional mass
observations such a discrimination could be possible if all the uncertainties
in the analysis can be accounted for.

\section{Estimating microphysical and astrophysical uncertainties}

To obtain robust results on whether a particular compositional scenario
can be realized in observed astrophysical sources, requires an assessment
of the many uncertainties in the process. The above expressions eqs.
(\ref{eq:luminosity-mass-accretion-relation}), (\ref{eq:non-accreted-envelope}),
(\ref{eq:accreted-envelope}), (\ref{eq:modified-Urca-emissivity}),
(\ref{eq:hadronic-Urca-emissivity}) and (\ref{eq:quark-Urca-emissivity})
depend both on bulk properties of astrophysical sources as well as
the microscopic uncertainties related to the underlying transport
and emission processes. In the following we estimate the uncertainties
of the various input parameters and determine their impact on the
relation eq. (\ref{eq:luminosity-mass-accretion-relation}) between
the quiescence thermal luminosity and the mass accretion rate. In
many cases it is surely impossible to rigorously determine the uncertainty
of poorly known astrophysical or microphysical input parameters to
a well defined significance. In these cases we estimate a \emph{conservative
maximum interval} that these parameters should, according to what
we know about the underlying physics, lie in. For a conservative analysis
we consider this ``maximum interval'' as a $2\sigma$ interval of
a Gaussian probability distribution, taking into account that what
is currently conceivable could be with a small probability too limited
or biased, and might have to be revised as new data or theoretical
insight becomes available. In the following we discuss all arising
parameters and estimate their ranges, and the resulting $1\sigma$
intervals are also compiled in table \ref{tab:paramter.ranges} individually:
\begin{description}
\item [{$M$}] For most considered astrophysical sources the neutron star
mass is not known, so that we have to estimate a mass range for these
sources. Theoretically the mass could be very low but it is limited
by the birth process in a core collapse supernova and our minimum
bound stems from the smallest observed neutron star mass, $M_{c}=\left(1.174\pm0.004\right)\,M_{\odot}$
obtained for the companion of the pulsar J0453+1559 \citep{Martinez2015}.
While these sources in double neutron star systems likely never accreted,
the sources in LMXBs studied here have an accretion history and should
likely be significantly heavier, but since it is hard to estimate
by how much we use $M_{\mathrm{min}}=1.2\,M_{\odot}$ as a conservative
lower bound for LMXBs as well. The maximum mass is harder to estimate,
but has to be larger than the masses of observed sources slightly
in excess of $2M_{\odot}$ \citep{Ozel:2016oaf}, including in particular
very precise measurements from Shapiro delay \citep{Demorest:2010bx,NANOGrav:2019jur}.
A more recent study \citep{Romani:2022jhd} found an even larger,
but somewhat more uncertain mass $M=\left(2.35\pm0.17\right)\,M_{\odot}$.
Actual upper bounds can be obtained from neutron star merger giving
a mass bound $M_{\mathrm{\mathrm{TOV}}}\lesssim\left(2.16\pm0.17\right)\,M_{\odot}$
\citep{Rezzolla:2017aly,Ruiz:2017due}. The masses of rotating stars
can be larger than the static TOV mass by $M_{\mathrm{rot}}\lesssim\left(1.20\pm0.05\right)\,M_{\mathrm{TOV}}$
\citep{Rezzolla:2017aly}. While for sources spinning with presently
observed spin frequencies the effect should be smaller, an exotic
composition might increase these values to some extent. When considering
these maximum masses one in particular has to take into account that
a theoretical equilibrium configuration with maximum TOV or rotational
mass (which will collapse at the smallest disturbance) is not a realistic
possibility for a source that is billions of years old and regularly
suffers cataclysmic events---like the accretion of huge amounts of
matter (up to nearly the mass of the moon in a short time interval
on an object the size of a city) or thermonuclear explosions covering
the entire surface of the star. Based on all these constraints we
choose a conservative upper bound of $2.5\,M_{\odot}$, resulting
in the $1\sigma$ range in table \ref{tab:paramter.ranges}.
\item [{$R$}] The radius is mainly determined by observations of the thermal
x-ray emission from the star's surface \citep{Lattimer:2013hma,Ozel:2015fia}.
Generally these studies have large uncertainties in mass-radius space,
but taking into account that radii from TOV solutions are generically
rather insensitive over a wide range of masses, the radius can be
constrained with sufficient statistics. The improved accuracy of NICER
allowed to obtain precise values for individual sources. The most
precise values stem from the $1.4\,\mathrm{M_{\odot}}$ star J0030+0451
\citep{Miller:2019cac} and the $\approx\!2.1\,M_{\odot}$ star J0740+6620
\citep{Miller:2021qha} which when combined with other constraints
yield expected radii $\left(12.45\pm0.65\right)\,\mathrm{km}$ for
$1.4\,\mathrm{M_{\odot}}$ and $\left(12.35\pm0.75\right)\,\mathrm{km}$
for $\approx\!2.1\,M_{\odot}$. Even heavier sources can be expected
to be more compact, in particular if they have an exotic composition
\citep{Alford:2013aca}. Based on all these studies we estimate the
maximum conservative interval for the radii of neutron stars in the
above mass range between $10\,\mathrm{km}$ and $13.7\,\mathrm{km}$.
This interval also includes the smaller impact of the non-catalyzed
crust on the equation of state and in turn on the structure of lighter
stars \citep{Fantina:2022xas}. 
\item [{$Q_{B}$}] Describing the deep crustal heating requires simulations
of the nuclear composition and of the various different reaction channels
as the accreted material is processed to heavier elements via electron
capture, neutron emission and in particular pycno-nuclear reactions
\citep{1990A&A...227..431H,Haensel2003,Haensel2008}. The pycno-nuclear
fusion reactions require tunneling and can only happen at sufficiently
high density. The total heating can be described by the average heat
energy per baryon $Q_{B}$ deposited within all these reactions deep
in the crust. Although the simulations are complicated and the detailed
reaction path can sensitively vary, the quantity $Q_{B}$ is fortunately
less sensitive since eventually ``the entire nuclear fuel is burned''
irrespective of the detailed reactions, and heats the star unless
part of it is lost via neutrino processes that are part of the reaction
path. To estimate these effects \citep{Haensel2008} considered two
extreme cases at which pycno-nuclear reactions start to operate ($A_{i}=56$
vs. $A_{i}=106$) as well as the extreme cases of maximum neutrino
losses and the complete blocking of the neutrino emission. This way
an interval $1.16\,\mathrm{MeV}\lesssim Q_{B}\lesssim1.93\,\mathrm{MeV}$
was obtained. Using a multi-component liquid drop model slightly
larger values where found in \citep{Steiner:2012bq}. More recently
the heating was considered in a density functional theory, considering
nuclear shell effects but neglecting neutrino losses \citep{Fantina:2018yad},
giving the narrow range $1.5\,\mathrm{MeV}\lesssim Q_{B}\lesssim1.7\,\mathrm{MeV}$.
In view of the still slightly differing results obtained in different
approximations we consider a conservative maximum interval of $1.2\,\mathrm{MeV}\lesssim Q_{B}\lesssim2\,\mathrm{MeV}$,
giving the $1\sigma$ uncertainty in table \ref{tab:paramter.ranges}.
\item [{$\tilde{L}_{i}$}] As discussed in section \ref{sec:Steady-state}
this parameter effectively takes into account the effect of gravity
on the neutrino emission, and describes its deviation in a realistic
star compared to a large uniform atomic nucleus. It depends both on
the equation of state and the particular stellar configuration (characterized
by its mass). Here we consider an APR equation of state \citep{Akmal:1998cf}
for hadronic matter and a bag model for quark matter and consider
different stellar configurations to numerically compute the constants
eq. (\ref{eq:luminosity-constant}) for the considered Urca processes.
We find that, aside for configurations close to the mass limit, the
impact of gravity is actually quite modest, while, as discussed, configurations
close to the mass limit are not realistic for these sources anyway.
Correspondingly considering APR configurations between $1.2\,M_{\odot}$
and $2\,M_{\odot}$ (while the maximum mass limit of the APR is already
at $2.2\,M_{\odot}$) yields e.g. for modified Urca reactions a range
$1.16\lesssim\tilde{L}_{\mathrm{mU}}\lesssim1.90$ and we double this
error range in order to take into account the uncertainties due to
the equation of state.
\item [{$\alpha_{\mathrm{Urca}}$}] The modified Urca amplitude has a non-power
law dependence that was originally absorbed in this factor in \citep{Friman1979}
which amounted to a small correction close to one. More recently in
\citep{Shternin:2018dcn} also off-shell effects, that were ignored
in the originally derivation \citep{Friman1979}, were included in
this factor and they result in a sizable enhancement of the rate and
according to the conventions of \citep{Shternin:2018dcn} also include
the gradual opening of the direct Urca channel \footnote{The integrals that define $\alpha_{\mathrm{Urca}}$ in this improved
approximation evaluate to simple analytic expressions describing its
mild density dependence away from threshold, but it eventually increases
dramatically over a narrow interval towards the 6-7 orders of magnitude
direct Urca emissivity. In this approximation $\alpha_{\mathrm{Urca}}$
even formally diverges at the Urca threshold, but this can easily
regulated to obtain a realistic behavior.}. In this work we consider the direct Urca channel separately and
in order to only take into account uncertainties due to the complicated
density dependence of the modified Urca rate we consider the behavior
sufficiently away from threshold, giving a conservative maximum interval
$5\leq\alpha_{\mathrm{Urca}}\leq13$.
\item [{$\beta_{\mathrm{Urca}}$}] Via this factor (called $R_{\mathrm{Urca}}$
in \citep{Friman1979}) that multiplies the (analytic) result for
the modified Urca rate due to long-range pion exchange \citep{Friman1979},
we take into account the effect of repulsive short-range correlations.
As a guide we considered the (model-dependent) results obtained in
the same article within an operator product expansion (OPE), which
increase the emissivity, and we further increased the uncertainty
range assuming a conservative maximum interval of $2/3\lesssim\beta_{\mathrm{Urca}}\lesssim2$.
\item [{$\alpha_{s}$}] Since the Urca rate vanishes for free, ultra-relativistic
particles \citep{Iwamoto1980,Alford:2011df}, neutrino emission in
dense quark matter requires strong interactions to open up phase space.
As discussed, we consider $\alpha_{s}$ as an effective parameter
that takes into account the effect of the non-perturbative strong
dynamics. There are no controlled non-perturbative QCD computations
at large baryon density so far \citep{Brambilla:2014jmp}. While the
coupling becomes small in the asymptotically free regime at large
scales, experimental studies and vacuum computations show that at
energies of a few hundred MeV, corresponding to the densities in neutron
stars, the coupling is still quite large and not so far from its infrared
value $\alpha_{s}\approx1-3$ \citep{Deur:2023dzc}. In order to consider
the impact of other non-perturbative effects we consider the upper
value also as the upper bound of our conservative interval for the
effective parameter $\alpha_{s}$ even in dense quark matter. Charge
screening should to some extent weaken the interaction, but the experimental
value for the running coupling from $\tau$-decay into hadrons takes
already a value $\alpha_{s}\left(m_{\tau}^{2}\right)\approx0.3$ at
the much larger scale $m_{\tau}\approx1.8\,\mathrm{GeV}$ \citep{ParticleDataGroup:2022pth}.
Effects due to charge screening and their impact on the quark self-energy
corrections, that affect their dispersion relations, only become relevant
at the significantly lower quark Fermi energy scales while both perturbative
and non-perturbative results predict that the coupling steeply rises
further below the $\tau$-scale. Therefore, expecting a perturbative
situation with a small coupling does not seem realistic and we choose
a conservative maximum interval $0.5\lesssim\alpha_{s}\lesssim3$
for the effective parameter $\alpha_{s}$ that takes into account
the combined effects of the strong (non-perturbative) dynamics. Our
lower bound is roughly consistent with the value chosen in \citep{Iwamoto1980}.
\item [{$m_{n/p}^{*}$}] In medium the effective baryon masses $m_{n/p}^{*}$
are expected to be somewhat reduced compared to their values in vacuum
due to many body effects and since the interactions that cause the
dynamic mass generation (as well as chiral symmetry breaking and confinement)
are already slightly reduced by both the running of the coupling and
charge screening. These three key phenomena of the strongly interacting
QCD vacuum are intimately linked, arising from the same non-perturbative
dynamics, and therefore it can be expected that in a confined phase
the in-medium baryon masses are not dramatically reduced. Correspondingly
we consider here a conservative maximum interval of $0.6\lesssim m_{n/p}^{*}/m_{n,p}\lesssim1$. 
\item [{$m_{s}$}] The fact that the strange quark is much heavier than
the light quarks enhances the Urca rate in quark matter. The in-medium
strange quark mass $m_{s}$ is poorly known due to the impossibility
to perform controlled computations at high density. Its lower limit
is the current quark mass arising from the Higgs mechanism $m_{s0}=93.4_{-3.4}^{+8.6}\,\mathrm{MeV}$
(in the $\overline{\mathrm{MS}}$ scheme at a renormalization scale
of $2\,\mathrm{GeV}$) \citep{ParticleDataGroup:2022pth}. Dynamical
mass generation reduces this value, but since confinement and dynamical
mass generation should be intimately linked, in a deconfined phase
we expect that the residual mass generation is limited and much smaller
than the constituent quark mass values in vacuum. Using a chiral NJL
model \citep{buballa2005njl},which lacks confinement and should therefore
overestimate the maximum mass of deconfined quarks, as an upper guide,
we estimate an upper value of $200\,\mathrm{MeV}$. We again consider
this as a conservative maximum interval, which gives the $1\sigma$
range for $m_{s}/m_{s0}$ in table \ref{tab:paramter.ranges}.
\item [{$x_{p}\!\left(n_{0}\right)$}] The method to absorb the entire
dependence of the neutrino emission on the equation of state and the
structure of the star in the single constant $\tilde{L}$, eq. (\ref{eq:luminosity-constant})
, requires in the hadronic case in addition merely the proton (or
lepton) fraction at saturation density $x_{p}\!\left(n_{0}\right)$.
Due to the chance to perform precise nuclear experiments this value
is much better constrained than the composition at high density.
We obtained its value from the analysis \citep{fraga20233} based
on chiral effective theory, which should work well in this regime
and yields the $1\sigma$ interval $x_{p}\!\left(n_{0}\right)=0.05\pm0.002$.
\end{description}
\begin{table}[H]
\begin{tabular}{ccc}
\hline 
Symbol & Explanation & Estimated value\tabularnewline
\hline 
\hline 
$Q_{B}$ & Average heat energy per baryon & $1.545^{\pm0.19}\,\text{MeV}$\tabularnewline
$M$ & Stellar mass & $1.85^{\pm0.33}\,M_{\odot}$\tabularnewline
$R$ & Stellar radius & $11.8^{\pm0.9}\,\mathrm{km}$\tabularnewline
$\tilde{L}_{mU}$ & Modified Urca luminosity constant & $1.41^{\pm0.24}$\tabularnewline
$\tilde{L}_{dU}$ & Direct Urca luminosity constant & $0.99^{\pm0.07}$\tabularnewline
$\tilde{L}_{q\beta}$ & Quark beta luminosity constant & $0.70^{\pm0.048}$\tabularnewline
$\alpha_{\mathrm{Urca}}$ & Urca off-shell correction factor  & $8^{\pm2.5}$\tabularnewline
$\beta_{\mathrm{Urca}}$ & Urca short-range correction & $0.83^{\pm0.22}$\tabularnewline
$x_{p}\!\left(n_{0}\right)$ & Electron fraction at saturation & $0.05^{\pm0.002}$\tabularnewline
$m_{n}^{*}/m_{n}$ & In-medium neutron mass fraction & $0.8^{\pm0.1}$\tabularnewline
$m_{p}^{*}/m_{p}$ & In-medium proton mass fraction & $0.8^{\pm0.1}$\tabularnewline
$\alpha_{s}$ & Strong coupling constant & $2^{\pm0.5}$\tabularnewline
$m_{s}^{*}/m_{s}$ & Strange quark mass fraction & $1.55^{\pm0.28}$\tabularnewline
\hline 
\end{tabular}

\caption{\label{tab:paramter.ranges}Estimated model parameters with 1$\sigma$
uncertainty interval, see the text for details.}
\end{table}

\section{Statistical analysis}

To estimate the compatibility of observed sources with different compositional
models, we perform a probability analysis. We assume that the various
uncertainties are independent and normally distributed. Performing
linear error propagation for a general function $f$ depending on
$n$ independent random variables $x_{i}$ with mean values $\bar{x}_{i}$
and standard deviations $\Delta x_{i}$ the mean value $\bar{f}$
and standard deviation $\Delta f$ of the function value is given
by \citep{ParticleDataGroup:2022pth}

\begin{align}
\bar{f} & =f\!\left(\bar{x}_{1},\cdots,\bar{x}_{n}\right)\nonumber \\
\Delta f & =\sqrt{\sum_{i=1}^{n}\left(\partial_{i}f\!\left(\bar{x}_{1},\cdots,\bar{x}_{n}\right)\right)^{2}\left(\Delta x_{i}\right)^{2}}\label{eq:error-propagation}
\end{align}
This expression is justified if the uncertainties are sufficiently
small $\Delta x_{i}/\bar{x}_{i}\ll1$. 

In reality the various input quantities are surely not completely
independent. For instance the radius and the mass of the star result
from the same solution of the Tolman-Oppenheimer-Volkov equation (depending
on the equation of state and the central pressure). However, in practice
for neutron stars in a wide range of realistic masses the radius is
very insensitive to the mass, which justifies the assumption in this
case. Similarly their poorly constrained probability distributions
are generally not perfectly symmetric normal distributions. For instance
neutron star masses have an upper and a lower cutoff. Assuming that
with a very small probability they can have masses outside of this
range we merely overestimate the uncertainty so that our estimates
are more conservative than necessary. In addition according to the
general version of the central limit theorem, the function $f$ will
in the large $n$ limit indeed be normally distributed, nearly irrespective
of the individual distributions of its various variables. In our case
with at most a dozen of dependent variables this limit is surely not
reached, but in practice it is found that a normal distribution can
be a good approximation already for moderate $n$. Therefore, we expect
that the assumption of $f$ being normally distributed with mean value
$\bar{f}$ and standard deviation $\Delta f$ is a reasonable approximation.
Likewise the observational X-ray data (stemming typically from a large
number of individual photon counts) can safely be assumed to be approximately
normally distributed.

Although we tried to estimate all key uncertainties in the analysis,
there are without a doubt further uncertainties that we have not explicitly
considered here. This concerns not the least the observational data
that can involve various systematic uncertainties and model assumptions.
In particular we do not consider errors for the average mass accretion
even though there are surely uncertainties involved. In this context
it is important to note that the overall error intervals eq. (\ref{eq:error-propagation})
considered here are quite conservative in the sense that we assumed
that all these errors add up. The more independent errors there are
this becomes more and more unlikely and the various errors should
rather partially cancel, which means that our present simplified statistical
analysis inherently overestimates the error intervals.

In order to assess the compatibility of a source with a particular
model for the cooling within the star, we compute the joint probability
of both the data ($D$) and the model ($M$). If they are normally
distributed, the two distribution functions $\pi\!\left(x;\bar{x}_{M},\Delta x_{M}\right)$
and $\pi\!\left(x;\bar{x}_{D},\Delta x_{D}\right)$ generally intersect
at two points $x_{1}<x_{2}$. The probability that a given model $M$
is consistent with a data point $D$ can be divided up in a sum of
the probabilities in the different regions and for independent uncertainties
of model and data this gives

\begin{align}
p\!\left(M\cap D\right) & =p\!\left(M,x\leq x_{1}\right)\times p\!\left(D,x\leq x_{1}\right)\nonumber \\
 & \qquad+p\!\left(M,x_{1}<x\leq x_{2}\right)\times p\!\left(D,x_{1}<x\leq x_{2}\right)\nonumber \\
 & \qquad+p\!\left(M,x_{2}<x\right)\times p\!\left(D,x_{2}<x\right)\label{eq:compatibility-probability}
\end{align}
For normal distributions the probabilities over the different intervals
are given in terms of the cumulative distribution function $F$

\begin{align*}
p\!\left(x\leq x_{1}\right) & =F\!\left(x_{1}\right)\\
p\!\left(x_{1}<x\leq x_{1}\right) & =F\!\left(x_{2}\right)-F\!\left(x_{1}\right)\\
p\!\left(x_{2}<x\right) & =1-F\!\left(x_{2}\right)
\end{align*}
 which in turn can be expressed via the error function
\[
F\!\left(x;\bar{x},\Delta x\right)\equiv\int_{-\infty}^{x}\!\!\pi\!\left(x^{\prime},\bar{x},\Delta x\right)dx^{\prime}=\frac{1}{2}\!\left(\!1\!+\!{\rm erf}\!\left(\frac{x-\bar{x}}{\sqrt{2}\Delta x}\right)\!\right)
\]
In the limit that data and model are clearly separated one can obtain
the analytic Gaussian expression 

\[
p\!\left(M\cap D\right)\xrightarrow[\Delta x_{1},\Delta x_{2}\ll\left|\bar{x}_{2}-\bar{x}_{1}\right|]{1}\sqrt{\frac{2}{\pi}}\frac{1}{\xi}\exp\left(-\frac{\xi^{2}}{2}\right)
\]
in terms of the single parameter

\[
\xi\equiv\frac{\bar{x}_{2}-\bar{x}_{1}}{\Delta x_{2}+\Delta x_{1}}
\]
which shows that the probability that they are consistent decreases
steeply when model and data are sufficiently constrained and far apart
(i.e. $\xi\gg1$). 

When it comes to the observational data, in our analysis we consider
these statistical expressions to take into account the more easily
quantifiable uncertainties in the luminosities in eq. (\ref{eq:luminosity-mass-accretion-relation}),
but eventually in the future a more realistic 2D statistical analysis
should be performed, that also takes into account the harder quantifiable
uncertainties in the average mass accretion rates of these sources. 

\section{Results}

Evaluating eq. (\ref{eq:luminosity-mass-accretion-relation}) using
the uncertainty ranges in tab. \ref{tab:paramter.ranges} and performing
linear error propagation according to eq. (\ref{eq:error-propagation})
we find asymptotic power law dependencies, where all uncertainties
are included in the uncertainty of the prefactors. In case of the
fast cooling mechanism these power laws are given for the case that
the entire star emits via these direct neutrino processes and they
present therefore the maximum possible cooling, while hybrid sources
where fast cooling is only possible in a smaller core would be somewhere
in between these extreme cases and will be discussed below. These
theoretical models are compared to data from various LMXBs taken from
table \citep{Potekhin2019,Potekhin:2023ets} and given in table \ref{tab:source-properties}. 

\begin{table}[t]
\begin{tabular}{cccc}
Source & $\langle\dot{M}_{obs}\rangle\,\left[M_{\odot}\mathrm{/y}\right]$ & $L_{q}\,\left[10^{33}\,\mathrm{erg/s}\right]$ & $M\,\left[M_{\odot}\right]$\tabularnewline
\hline 
\hline 
4U 2129+47 & $3.9\times10^{-9}$ & $1.5_{-1.2}^{+3.1}$ & \tabularnewline
KS 1731\textminus 260 & $<9\times10^{-10}$ & $0.39^{\pm0.03}$ & $1.61\pm0.37$\tabularnewline
4U 1608\textminus 522 & $9.6\times10^{-10}$ & $5.3_{-2.9}^{+4.7}$ & $1.57\pm0.30$\tabularnewline
1M 1716\textminus 315 & $<2.5\times10^{-10}$ & $1.3_{-0.7}^{+1.2}$ & \tabularnewline
XTE J1709 & $1.8\times10^{-10}$ & $1.4_{-0.5}^{+0.6}$ & \tabularnewline
MXB 1659\textminus 29 & $1.4\times10^{-10}$ & $0.2_{-0.11}^{+0.05}$ & \tabularnewline
Cen X-4 & $3.8\times10^{-11}$ & $0.12^{\pm0.01}$ & \tabularnewline
4U 1730\textminus 22 & $<4.8\times10^{-11}$ & $2.2_{-1.1}^{+2.0}$ & \tabularnewline
Aql X-1 & $3.2\times10^{-10}$ & $2.1^{\pm0.5}$ & \tabularnewline
NGC 6440 X-1 & $6\times10^{-11}$ & $1.3^{\pm0.4}$ & \tabularnewline
IGR J00291 & $\sim2.2\times10^{-12}$ & $0.19_{-0.08}^{+0.06}$ & \tabularnewline
HETE J1900.1 & $3.9\times10^{-11}$ & $0.061^{\pm0.037}$ & \tabularnewline
Ter5 X-2 & $<1.7\times10^{-11}$ & $0.7^{\pm0.1}$ & \tabularnewline
EXO 0748 & $<4.4\times10^{-10}$ & $3.8^{\pm0.2}$ & \tabularnewline
1RXS J180408 & $<4.6\times10^{-11}$ & $0.74_{-0.18}^{+0.09}$ & \tabularnewline
Ter5 X-3 & $<3\times10^{-11}$ & $1.2^{\pm0.2}$ & \tabularnewline
\hline 
\end{tabular} \caption{\label{tab:source-properties}Accretion observables for transiently
accreting LMXBs taken from \citep{Potekhin2019}, that have an actual
quiescence luminosity measurement (in contrast to only a bound), extended
by mass measurements taken from \citep{Ozel:2016oaf} where available.}
\end{table}

\begin{figure*}
\includegraphics[scale=0.6]{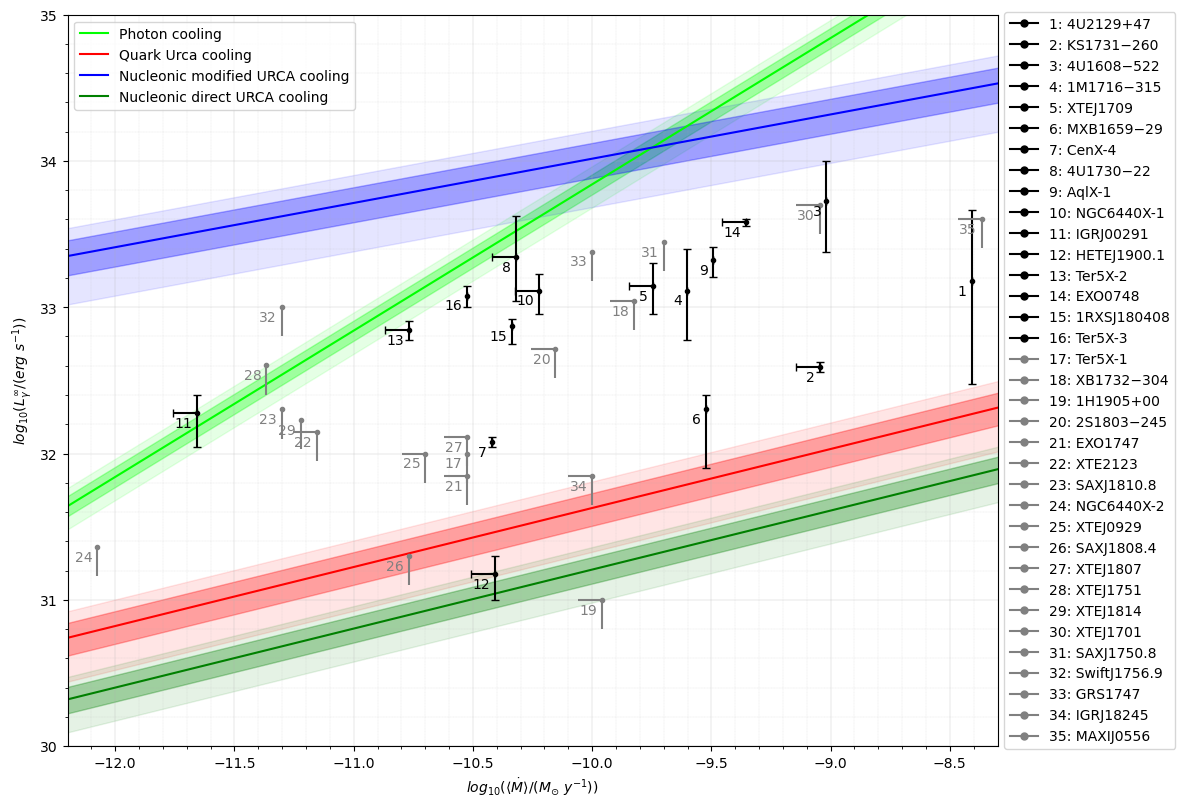}\caption{\label{fig:fully-accreted-envelope}Red-shifted quiescence photon
luminosity $L_{\gamma}^{\infty}$ of sources with different dominant
emission mechanisms versus averaged accretion rate $\left\langle \dot{M}\right\rangle $
for a fully accreted envelope model \citep{Potekhin1997} compared
to LMXB data \citep{Potekhin2019,Potekhin:2023ets}. Dark and bright
colored bands around the theoretical model curves show $1\sigma$
and $2\sigma$ errors respectively,. The observational errors for
the sources in black are $1\sigma$ intervals, while sources in grey
only offer upper bounds.}
\end{figure*}

\begin{figure*}
\includegraphics[scale=0.6]{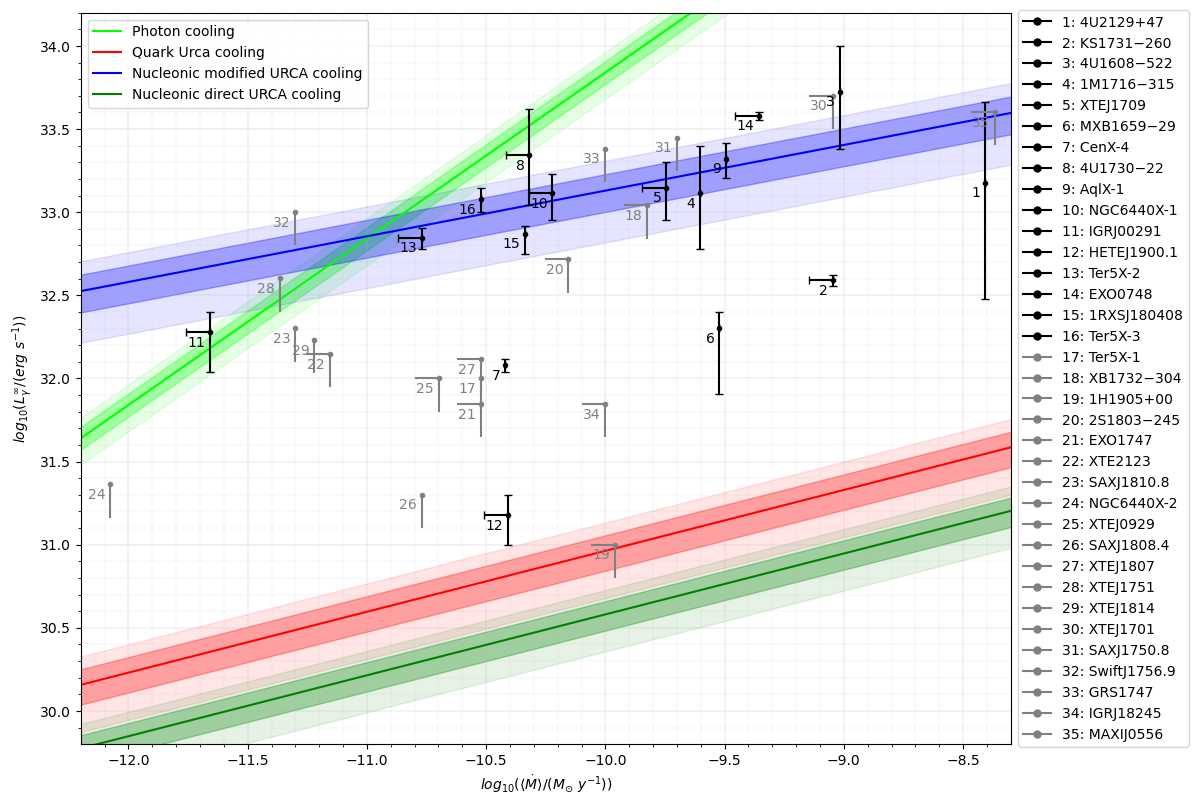}\caption{\label{fig:non-accreted-envelope}Red-shifted quiescence photon luminosity
$L_{\gamma}^{\infty}$ of sources with different dominant emission
mechanisms versus averaged accretion rate $\left\langle \dot{M}\right\rangle $
for a fully-catalyzed (non-accreted) iron envelope model \citep{Gudmundsson1982}
compared to LMXB data \citep{Potekhin2019,Potekhin:2023ets}. Dark
and bright colored bands around the theoretical model curves show
$1\sigma$ and $2\sigma$ errors respectively,. The observational
errors for the sources in black are $1\sigma$ intervals, while sources
in grey only offer upper bounds.}
\end{figure*}

The results are shown in fig. \ref{fig:fully-accreted-envelope} for
a non-accreted envelope and in fig. \ref{fig:non-accreted-envelope}
for a fully-accreted envelope, respectively. As can be seen, despite
the substantial uncertainties the curves are in both cases well separated
at the $1\sigma$-level and do not significantly overlap even at the
$2\sigma$-level. This is mainly due to the mild parametric dependence
on most parameters in eq. (\ref{eq:luminosity-mass-accretion-relation}).
Therefore, the different models are, even within the various uncertainties,
in principle distinguishable. 

The minimal cooling of a neutron star is due to slow hadronic modified
Urca reactions \footnote{There are slower processes like bremsstrahlung \citep{Yakovlev2001}
that can be present in addition, but modified Urca should be realized
unless large parts of the star are superfluid and superconducting,
which is not expected. Therefore, it is rather the fastest of the
slow processes that are expected for a minimal (npe$\mu$) neutron
star composition, while various enhanced processes are only present
under certain conditions.} and in this case photon cooling from the surface dominates at low
accretion rates. The transition is around $10^{-11}\,M_{\odot}/\mathrm{y}$
for a non-accreted envelope and above $10^{-10}\,M_{\odot}/\mathrm{y}$
for a fully accreted envelope. In the absence of enhanced cooling
LMXBs accreting below the lower value inherently cannot probe the
interior composition and in the considered data set there are only
two such sources (IGR J00291 \& XTE J1751) which are well consistent
with surface photon emission, while others seem to require enhanced
cooling. What is more, for a catalyzed envelope also nearly all hotter
sources are well compatible with the minimal modified Urca model.
While it is interesting that in the case of a fully accreted envelope
fig. \ref{fig:fully-accreted-envelope} photon emission can within
uncertainties likewise explain the hotter weakly accreting sources,
the strongly accreting sources are in this case even within uncertainties
hardly compatible with the minimal modified Urca model and would require
a stronger emission mechanism, like pair-breaking emission in nuclear
superfluids \citep{Yakovlev2001}. 

In reality the envelope composition varies over time and should be
somewhere in between these two extreme cases. Even the average composition
could furthermore in principle vary from source to source depending
on both the average accretion rate and the number of outbursts. However,
as seen from figs. \ref{fig:non-accreted-envelope} and \ref{fig:fully-accreted-envelope},
an intermediate envelope with only limited light element composition
(i.e. closer to fig. \ref{fig:fully-accreted-envelope}) is within
uncertainties consistent with all hotter sources. This might to some
extent be surprising since higher accretion rates might suggest a
larger fraction of light elements. Therefore, this would require an
efficient burning of light elements in these sources. Since these
sources don't show any bursts or other signs of appreciable accretion
in quiescence, a potential explanation might be that when accretion
stops the accreted light elements eventually explosively burn without
leaving large leftovers and this marks the end of the outburst phase.
Such a scenario would not require superfluidity for most sources and
taking into account that pair-breaking emission typically only stems
from particular density ranges at the boundaries of the superfluid
region, this scenario would be far less sensitive to the poorly constrained
density dependence of superfluid gaps \citep{Sedrakian:2018ydt}.

Irrespective of the envelope composition, there are several sources
that even within the sizable uncertainties significantly deviate from
the minimal cooling scenario and clearly require enhanced cooling.
In particular, the coldest of these sources seem to require some fast
cooling mechanism involving direct Urca reactions. In case of a fully
accreted envelope only a maximal core with hadronic direct Urca \citep{Lattimer1994}---being
the most efficient known cooling mechanism \footnote{Hyperonic direct Urca could have a slightly higher emissivity than
purely non-strange reactions.\citep{Yakovlev2004}.}---seems to be able to provide the required strength to explain the
coldest source 1H 1905+00 \footnote{Assuming the current bound for this faint source is not too low as
was the case for SAX 1808.4 and which had to be corrected upwards
quite substantially over time \citep{Han2017,Potekhin2019}.} while quark Urca emission in a sufficiently large core could potentially
explain several of the colder sources (SAX 1808.4, HETE J1900.1, ...).

\begin{table*}[t]
\begin{tabular}{cc|ccccccccc|cccc}
Source &  &  & $p_{\gamma}$ &  & $p_{\mathrm{mU}}$ &  & $p_{\mathrm{dU}}$ &  & $p_{\mathrm{qU}}$ &  &  & $M_{c,\mathrm{dU}}/M$ &  & $M_{c,\mathrm{qU}}/M$\tabularnewline
\hline 
\hline 
4U 2129+47 &  &  & - &  & $8.5\times10^{-3}$ &  & $1.2\times10^{-2}$ &  & $3.4\times10^{-2}$ &  &  & $5.1\times10^{-4}$ &  & $5.9\times10^{-3}$\tabularnewline
KS 1731\textminus 260$^{*}$ &  &  & - &  & $2.8\times10^{-5}$ &  & $-$ &  & $2.8\times10^{-7}$ &  &  & $3.3\times10^{-3}$ &  & $3.8\times10^{-2}$\tabularnewline
4U 1608\textminus 522 &  &  & - &  & $1.3\times10^{-1}$ &  & $2.9\times10^{-3}$ &  & $8.7\times10^{-3}$ &  &  & $5.4\times10^{-6}$ &  & $6.3\times10^{-5}$\tabularnewline
1M 1716\textminus 315$^{*}$ &  &  & - &  & $9.6\times10^{-3}$ &  & $6.5\times10^{-3}$ &  & $2.0\times10^{-2}$ &  &  & $4.7\times10^{-5}$ &  & $5.4\times10^{-4}$\tabularnewline
XTE J1709 &  &  & - &  & $3.9\times10^{-3}$ &  & $1.6\times10^{-3}$ &  & $5.4\times10^{-3}$ &  &  & $2.8\times10^{-5}$ &  & $3.2\times10^{-4}$\tabularnewline
MXB 1659\textminus 29 &  &  & - &  & $7.8\times10^{-5}$ &  & $4.3\times10^{-4}$ &  & $1.2\times10^{-2}$ &  &  & $2.7\times10^{-3}$ &  & $3.1\times10^{-2}$\tabularnewline
Cen X-4 &  &  & - &  & $2.4\times10^{-5}$ &  & $-$ &  & $1.1\times10^{-7}$ &  &  & $2.6\times10^{-3}$ &  & $3.0\times10^{-2}$\tabularnewline
4U 1730\textminus 22$^{*}$ &  &  & $6\times10^{-1}$ &  & - &  & $2.1\times10^{-3}$ &  & $6.4\times10^{-3}$ &  &  & $2.4\times10^{-6}$ &  & $2.8\times10^{-5}$\tabularnewline
Aql X-1 &  &  & - &  & $3.1\times10^{-3}$ &  & $8.2\times10^{-6}$ &  & $3.7\times10^{-5}$ &  &  & $1.8\times10^{-5}$ &  & $2.1\times10^{-4}$\tabularnewline
NGC 6440 X-1 &  &  & $4.2\times10^{-2}$ &  & - &  & $1.4\times10^{-4}$ &  & $5.1\times10^{-4}$ &  &  & $1.1\times10^{-5}$ &  & $1.3\times10^{-4}$\tabularnewline
IGR J00291 &  &  & $1.3\times10^{-1}$ &  & - &  & $3.3\times10^{-4}$ &  & $1.4\times10^{-3}$ &  &  & $4.8\times10^{-5}$ &  & $5.6\times10^{-4}$\tabularnewline
HETEJ1900.1 &  &  & - &  & $1.0\times10^{-5}$ &  & $3.9\times10^{-1}$ &  & $2.2\times10^{-1}$ &  &  & $4.6\times10^{-1}$ &  & -\tabularnewline
Ter5 X-2$^{*}$ &  &  & $1.8\times10^{-1}$ &  & - &  & $4.3\times10^{-12}$ &  & $6.5\times10^{-11}$ &  &  & $1.5\times10^{-5}$ &  & $1.7\times10^{-4}$\tabularnewline
EXO 0748$^{*}$ &  &  & - &  & $2.6\times10^{-3}$ &  & $-$ &  & $-$ &  &  & $5.7\times10^{-6}$ &  & $6.6\times10^{-5}$\tabularnewline
1RXS J180408$^{*}$ &  &  & $1.6\times10^{-2}$ &  & - &  & $1.7\times10^{-15}$ &  & $2.8\times10^{-13}$ &  &  & $3.5\times10^{-5}$ &  & $4.0\times10^{-4}$\tabularnewline
Ter5 X-3$^{*}$ &  &  & $2.6\times10^{-1}$ &  & - &  & $1.2\times10^{-9}$ &  & $7.0\times10^{-9}$ &  &  & $6.8\times10^{-6}$ &  & $7.9\times10^{-5}$\tabularnewline
\hline 
\end{tabular} \caption{\label{tab:fully-accreted-envelope}Probabilities for the compatibility
of different core cooling models ($\gamma$: photon emission, mU:
nucleonic modified Urca, dU: nucleonic direct Urca, qU: quark Urca)
with the various sources as well as hybrid star core mass fractions
for a fully-accreted light element envelope \citep{Potekhin1997}.
Sources marked by a star only have an upper bound on the mass accretion
rate.}
\end{table*}

\begin{table*}[t]
\begin{tabular}{cc|ccccccccc|cccc}
Source &  &  & $p_{\gamma}$ &  & $p_{\mathrm{mU}}$ &  & $p_{\mathrm{dU}}$ &  & $p_{\mathrm{qU}}$ &  &  & $M_{c,\mathrm{dU}}/M$ &  & $M_{c,\mathrm{qU}}/M$\tabularnewline
\hline 
\hline 
4U 2129+47 &  &  & $-$ &  & $3.3\times10^{-1}$ &  & $2.6\times10^{-3}$ &  & $7.0\times10^{-3}$ &  &  & $3.1\times10^{-6}$ &  & $3.4\times10^{-5}$\tabularnewline
KS 1731\textminus 260$^{*}$ &  &  & $-$ &  & $7.6\times10^{-4}$ &  & $-$ &  & $-$ &  &  & $2.9\times10^{-5}$ &  & $3.2\times10^{-4}$\tabularnewline
4U 1608\textminus 522 &  &  & $-$ &  & $2.0\times10^{-1}$ &  & $6.7\times10^{-4}$ &  & $1.8\times10^{-3}$ &  &  & $-$ &  & $-$\tabularnewline
1M 1716\textminus 315$^{*}$ &  &  & $-$ &  & $4.0\times10^{-1}$ &  & $1.6\times10^{-3}$ &  & $4.3\times10^{-3}$ &  &  & $2.0\times10^{-7}$ &  & $2.2\times10^{-6}$\tabularnewline
XTE J1709 &  &  & $-$ &  & $5.0\times10^{-1}$ &  & $3.7\times10^{-4}$ &  & $1.0\times10^{-3}$ &  &  & $6.4\times10^{-8}$ &  & $7.1\times10^{-7}$\tabularnewline
MXB 1659\textminus 29 &  &  & $-$ &  & $2.0\times10^{-3}$ &  & $3.3\times10^{-5}$ &  & $1.5\times10^{-4}$ &  &  & $2.8\times10^{-5}$ &  & $3.1\times10^{-4}$\tabularnewline
Cen X-4 &  &  & $-$ &  & $3.6\times10^{-4}$ &  & $-$ &  & $-$ &  &  & $3.1\times10^{-5}$ &  & $3.4\times10^{-4}$\tabularnewline
4U 1730\textminus 22$^{*}$ &  &  & $-$ &  & $2.1\times10^{-1}$ &  & $5.5\times10^{-4}$ &  & $1.5\times10^{-3}$ &  &  & $-$ &  & $-$\tabularnewline
Aql X-1 &  &  & $-$ &  & $4.8\times10^{-1}$ &  & $1.6\times10^{-6}$ &  & $4.8\times10^{-6}$ &  &  & $-$ &  & $-$\tabularnewline
NGC 6440 X-1 &  &  & $-$ &  & $5.2\times10^{-1}$ &  & $3.3\times10^{-5}$ &  & $9.3\times10^{-5}$ &  &  & $-$ &  & $-$\tabularnewline
IGR J00291 &  &  & $1.3\times10^{-1}$ &  & $-$ &  & $8.3\times10^{-5}$ &  & $2.4\times10^{-4}$ &  &  & $4.9\times10^{-7}$ &  & $5.4\times10^{-6}$\tabularnewline
HETE J1900.1 &  &  & $-$ &  & $4.5\times10^{-5}$ &  & $1.5\times10^{-2}$ &  & $1.5\times10^{-1}$ &  &  & $9.2\times10^{-3}$ &  & $1.0\times10^{-1}$\tabularnewline
Ter5 X-2$^{*}$ &  &  & $-$ &  & $4.3\times10^{-1}$ &  & $6.6\times10^{-13}$ &  & $2.4\times10^{-12}$ &  &  & $5.1\times10^{-8}$ &  & $5.7\times10^{-7}$\tabularnewline
EXO 0748$^{*}$ &  &  & $-$ &  & $1.2\times10^{-2}$ &  & $-$ &  & $-$ &  &  & $-$ &  & $-$\tabularnewline
1RXS J180408$^{*}$ &  &  & $-$ &  & $2.4\times10^{-1}$ &  & $1.1\times10^{-16}$ &  & $6.1\times10^{-16}$ &  &  & $2.0\times10^{-7}$ &  & $2.2\times10^{-6}$\tabularnewline
Ter5 X-3$^{*}$ &  &  & $-$ &  & $4.3\times10^{-1}$ &  & $2.3\times10^{-10}$ &  & $7.1\times10^{-10}$ &  &  & $-$ &  & $-$\tabularnewline
\hline 
\end{tabular}\caption{\label{tab:non-accreted-envelope}Probabilities for the compatibility
of different core cooling models with the various sources as well
as hybrid star core mass fractions for a catalyzed iron envelope \citep{Gudmundsson1982}.
Sources marked by a star only have an upper bound on the mass accretion
rate.}
\end{table*}

To quantify the above qualitative statements we consider the probabilities
eq. (\ref{eq:compatibility-probability}) for the compatibility of
a particular source with one of the considered models, which are given
in tabs. \ref{tab:fully-accreted-envelope} and \ref{tab:non-accreted-envelope},
respectively. Unfortunately in our statistical analysis we could not
consider some of the coldest sources, since they only have an upper
luminosity bound and so our simplified Gaussian analysis does not
apply (such sources are drawn in grey in figs. \ref{fig:non-accreted-envelope}
and \ref{fig:fully-accreted-envelope}). However, there are nonetheless
several sources with rather low luminosity measurements, and those
with small observational uncertainties are particularly promising.
Irrespective of the envelope model the three sources KS1731\textminus 260,
Cen X-4 and HETE J1900.1 are inconsistent with the minimal cooling
model at the $3.5\sigma$ level, and in case of a fully accreted envelope
even at the $4\sigma$ level. In addition MXB 1659\textminus 29 challenges
the minimal model at the $3\sigma$ level, so that the combined set
of all these sources is clearly statistically incompatible with the
minimal model. This shows quantitatively that even within the sizable
uncertainties enhanced cooling is required to explain the colder sources
in the LMXB data set.

Considering now the compatibility of the extended fast cooling mechanisms
with the data, although not surprising, there are several of the hotter
sources where fast cooling can be clearly ruled out at better than
the $5\sigma$ level. In case of a non-accreted crust, where the hotter
sources are well described by modified Urca emission, those that are
not, are also not well compatible with either of the extreme fast
cooling models. Therefore either some intermediate mechanism \citep{Yakovlev2001}
(e.g. in superfluids \citep{Potekhin2019}) or a hybrid source with
a smaller inner core, either made of quark matter or where hadronic
Urca reactions are allowed, would be required to explain these sources,
as discussed below. In contrast for a fully accreted crust even the
extreme quark matter model \footnote{In the case of quark matter such a strange star with a maximal quark
core would likely require the strange matter hypothesis \citep{Witten:1984rs}.} is marginally consistent with some of the colder sources. In this
context it is important to note that the quark Urca emissivity can
be enhanced by non-Fermi liquid effects \citep{Schafer:2004jp} stemming
from unscreened interactions in dense quark matter. These introduce
logarithmic temperature dependences (eq. (\ref{eq:emissivity-parametrization}))
that can be sizable at very low temperature, but are expected to be
mild for these heated sources. Including these effects in the future
will require a computationally somewhat more elaborate analysis. 

When it comes to the actual compatibility of a data point with a given
model the larger probabilities in the tables are less conclusive.
The reason for this is that even if model and data agree well with
each other the computed probability can depend strongly on the uncertainties.
For instance, for a data point with a very small uncertainty and a
model centered at exactly the same point the computed probability
is the smaller the larger the model uncertainties are, so that merely
our ignorance, or aspiration to estimate a conservative interval,
can make the probability small. In contrast, if model and data are
far apart a larger uncertainty actually increases the probability
that they are compatible. Therefore, if we nonetheless find a small
probability for the compatibility this represents an upper bound even
if we overestimated the uncertainties. Correspondingly, sufficiently
large probabilities in the tables merely qualitatively signal compatibility,
but cannot really quantify it, whereas small probabilities indeed
quantify the incompatibility.

Finally we estimate the sizes of the required fast cooling cores to
explain the observational sources with either of the fast cooling
models. Using eq. (\ref{eq:mass-fraction}), yields the mass fractions
given in tables \ref{tab:fully-accreted-envelope} and \ref{tab:non-accreted-envelope},
where our present analysis does not allow us to consistently consider
the uncertainties in the analysis. As can be seen for many of the
hotter sources the required cores are small for quark matter and even
significantly smaller for hadronic Urca cooling. Yet, for several
of the colder sources the required hybrid star cores are larger, and
when taking into account the uncertainties in the analysis it can
be expected that even larger quark cores than in tables \ref{tab:fully-accreted-envelope}
and \ref{tab:non-accreted-envelope} are compatible with the data.
Furthermore if quark matter is color superconducting \citep{Alford:2007xm}
the emissivity is reduced---e.g. in the classic 2SC phase by an effective
reduction factor 5/9 reflecting the ungapped quark flavors---bringing
such a model even closer to the data.

\section{Conclusion}

In this work we assessed the deep crustal heating scenario, taking
into account the various micro- and astrophysical uncertainties in
the analysis. We find that despite the numerous, and in several cases
sizable, uncertainties, deep crustal heating has the potential to
probe the properties of the star's interior and discriminate different
compositional scenarios. This is not the least due to the fortunate
fact that the relation between the astrophysical observables eq. (\ref{eq:luminosity-mass-accretion-relation})
is, similarly to the case of r-mode damping \citep{Alford:2010fd,Alford:2013pma},
rather insensitive to key uncertainties in the analysis. 

As discussed, the composition of the crust has a strong influence
on the cooling. We find that a mostly catalyzed envelope can consistently
explain all the hotter sources via standard modified Urca emission,
while assuming a strongly accreted envelope would require enhanced
cooling e.g. due to superfluidity. Observed sources should be in between
the studied extreme cases. This would mean that if modified Urca is
indeed the dominant mechanism in these sources, the most outer layers
of the crust responsible for its heat insulation should catalyze quickly
so that only little light elements remain during most of the quiescence
period. In this context it is interesting that the relaxation of the
crust observed by the time-dependent cooling behavior during quiescence
\citep{Brown:2009kw} can potentially directly probe the composition
of the crust. This would eliminate this key uncertainty and thereby
enable a more rigorous analysis to probe the core composition.

However, our analysis also shows that even when taking into account
these various uncertainties several colder sources clearly statistically
challenge the minimal scenario where LMXBs merely cool via slow modified
Urca reactions with an emissivity $\sim\left(T/p_{F}\right)^{8}$.
Therefore, it highlights the well known conclusion that there is very
likely not a single mechanism that can explain all observed sources
\citep{Yakovlev2004,Wijnands2017,Potekhin2019,Potekhin:2023ets},
pointing to some sort of extended composition beyond minimal $npe\mu$-matter. 

In this work we considered selected extended models of the interior
composition of neutron stars, featuring fast neutrino emission processes
$\sim\left(T/p_{F}\right)^{6}$, namely hadronic direct Urca and quark
Urca. In particular we considered here the extreme case that the entire
star emits via a single mechanism and find that such an strong cooling
is statistically inconsistent with all but the coldest sources. Yet,
if fast cooling is only realized in an inner core whose size depends
on details of the equation of state this naturally reduces the emissivity.
There are also intermediate processes $\sim\left(T/p_{F}\right)^{7}$
\citep{Yakovlev:2000jp}, e.g. from pair-breaking and formation in
superfluids or meson condensates, that have the potential to explain
the data and it would be interesting to perform an analogous statistical
analysis for such compositions in the future. When it comes to superfluidity
\citep{Sedrakian:2018ydt}, generically only part of the star can
be superfluid and it depends strongly on the density dependence of
the superfluid gap. Moreover, there are various other exotic phases
of matter that could be compared to the data to see if these are consistent
with a subset of the sources. 

\begin{figure}
\includegraphics[scale=0.58]{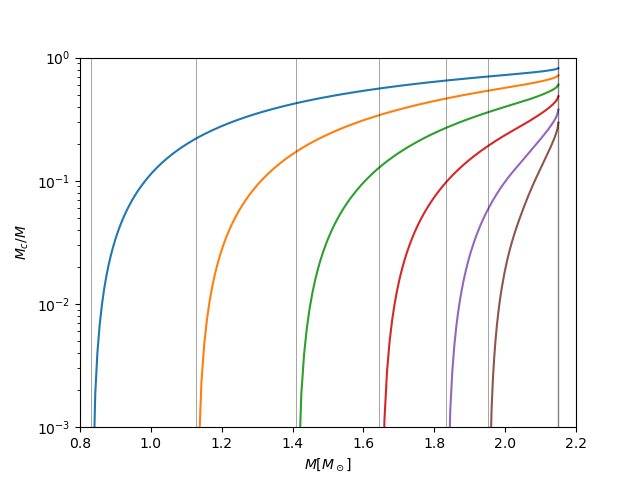}\caption{\label{fig:hybrid-core}Mass ratio $M_{c}/M$ of an inner core (with
enhanced neutrino emission) for various exemplary values of the transition
density as a function of the mass $M$ of the star. Thin vertical
lines mark the corresponding critical masses where an inner core appears
and the thick vertical line marks the maximum mass configuration.}
\end{figure}

Taking into account that a minimal hadronic (npe$\mu$) composition
can feature at the same time both the slowest \citep{Friman1979}
and nearly the fastest known cooling processes \citep{lattimer1991direct}
and that the cooling luminosity can depending on the not directly
observable size of the direct Urca core in principle even take any
intermediate value, it might seem that it is inherently impossible
to detect any exotic composition through its cooling behavior in transiently
accreting neutron stars. And even more so, when considering that the
probably most interesting exotic composition---deconfined quark matter---likewise
can based on the size of the quark core yield in principle any intermediate
cooling behavior. However, our results here show that the situation
is actually much more promising. To see this recall that according
to figs. \ref{fig:fully-accreted-envelope} and \ref{fig:non-accreted-envelope}
(maximal) quark Urca and hadronic direct Urca cooling are, even within
the sizable uncertainties, clearly distinguishable. The quark matter
and hadronic Urca emissivities are many orders of magnitude larger
than modified Urca. Using eq. (\ref{eq:mass-fraction}), this would
require dramatically small cores to explain these sources, as seen
in tables \ref{tab:fully-accreted-envelope} and \ref{tab:non-accreted-envelope}.
However, as seen in fig. \ref{fig:hybrid-core}, a very small quark
core would require a star mass just above the critical value for a
given equation of state where an inner core arises. Since the masses
stem from the accretion history and should be continuously distributed,
such a clustering of several sources in a very narrow interval would
be extremely unlikely and therefore generally only sufficiently large
cores are expected. Correspondingly, with sufficient data such a scenario
could be statistically challenged. Moreover, with additional mass
measurements for these sources such a scenario can be directly checked
since the source mass would have to monotonously increase from sources
with large to small photon emissivity (ranging from slow to fast neutrino
cooling).

Interestingly, there are already mass estimates for the two sources
KS 1731\textminus 260 and 4U 1608\textminus 522 \citep{Ozel:2016oaf}
which might have similar mass accretion rates (the value for the former
source being only an upper bound) and also have very similar mass
estimates. Yet, they have markedly different quiescence luminosities
that differ by more than an order of magnitude, so that 4U 1608\textminus 522
is consistent with modified Urca cooling while KS 1731\textminus 260
could based on the envelope composition be compatible with being a
hybrid star. The colder of them KS 1731\textminus 260 indeed seems
to have a slightly larger mass (although this is so far not significant
within the large uncertainties), so that if this mass ordering persists
in more precise measurements and quark matter would indeed be the
reason for its low luminosity, this could point to a critical mass
where quark matter arises around $1.6\,M_{\odot}$.

Taking into account that, even within uncertainties, hadronic and
quark phases of dense matter can show a markedly different fast cooling
behavior, there is a realistic chance that nucleonic and exotic phases
can be discriminated via a careful statistical analysis based on a
sufficiently large and precise astrophysical dataset. Such a discrimination
of different compositional scenarios of the neuron star interior is
even more promising considering other observed dynamic processes that
can probe the cooling behavior or even other material properties.
A prime example is the damping of r-modes both in LMXBs and other
sources, which also probes the viscous dissipation and thereby imposes
independent constraints that can distinguish different phases of matter
\citep{Andersson:2000mf,Alford:2010fd,Alford:2013pma}.
\begin{acknowledgments}
This work was supported by the Turkish Research Council (TÜBITAK)
via project 117F312.
\end{acknowledgments}

\bibliographystyle{apsrev4-1}
\bibliography{bibtex-thesis-library}

\end{document}